# Advances in Computational Biology:
# A Real Boost or a Wishful Thinking

Emanuel Diamant
*VIDIA-mant, Kiriat Ono, Israel*
emanl.245@gmail.com

**Abstract**
Computational biology is on the verge of a paradigm shift in its research practice – from a data-based (computational) paradigm to an information-based (cognitive) paradigm. As in the other research fields, this transition is impeded by lack of a right understanding about what is actually hidden behind the term "information". The paper is intended to clarify this issue and introduces two new notions of "physical information" and "semantic information", which together constitute the term "information". Some implications of this introduction are considered.

## 1  Introduction

Striking advances in high-throughput sequencing technologies have resulted in a tremendous increase in the amounts of data related to various biological screening experiments. Consequently, that gave rise to an urgent need of new techniques and algorithms for analyzing, modeling and interpreting these amounts of data.

This is announced as the main purpose of Computational biology – to organize, and to make sense of these amounts of data… and to use this information to make predictions and obtain experimentally testable models, [5].

To reach its goals, Computational biology involves the use of computational tools to discover new information in complex data sets and to decipher the languages of biology (e.g., the one-dimensional information of DNA, the three-dimensional information of proteins, and the four-dimensional information of living systems), [2].

As a theoretical foundation for a suitable model of the information flow that runs at each level and through all levels of biological organization the most often used theoretical model is the Shannon's Information theory. In [11], a thorough analysis of various aspects of its application to tasks of computational biology is provided.

Information Theory was at first introduced in Shannon's seminal paper "A Mathematical Theory of Communication" in 1948, [6]. The original aim of the theory was to solve a purely technical problem: to increase the performance of communication systems. In the theory, messages are characterized only by their probabilities, regardless of their value or meaning. Shannon was aware about this peculiarity of his theory. In the year 1949, he wrote: "These semantic aspects of communication are irrelevant to the engineering problem… It is important to emphasize, at the start, that we are not concerned with the meaning or the truth of messages; semantics lies outside the scope of mathematical information theory", [7].

However, the scientific community, fascinated with the new theory, was eager to apply it in a wide range of research areas. That forced Shannon to issue an additional warning (in 1956): "In short, information theory is currently partaking of a somewhat heady draught of general popularity. It will be all too easy for our somewhat artificial prosperity to collapse overnight when it is realized that the use of a few exciting words like information, entropy, redundancy, do not solve all our problems", [8].

In biology, semantic aspects of messages are of a paramount importance. However, the mainstream of computational biology usually ignore the Shannon's warnings. Even today, the matters of interrelation between the notion of "information" and the notions of "data", "knowledge", and "semantics" are still undefined, blurred and intuitive (due to the heritage of Information Theory).

It must be mentioned (in this regard) that the first attempt to clarify the relations between "information" and "semantics" was made about 60 years ago by Yehoshua Bar-Hillel and Rudolf Carnap, [1]. As to my knowledge, they were the first who coined the term "Semantic Information". They have sincerely believed that such a merging can be possible: "Prevailing theory of communication (or transmission of information) deliberately neglects the semantic aspects of communication, i. e., the meaning of the messages… Instead of dealing with the information carried by letters, sound waves, and the like, we may talk about the information carried by the sentence", [1].

However, they were not successful in their attempt to unite the mathematical theory of information and semantics. The mainstream thinking at that time was determined by The Mathematical Theory of Communication. Today, Stanford Encyclopedia of Philosophy distinguishes (beside Shannon's definition of information) a long list of other attempts to formalize the concept of information: Fisher information, Kolmogorov complexity, Quantum Information, Information as a state of an agent, and Semantic Information (once developed by Bar-Hillel and Carnap), [8].

None of the above listed proposals does not resolve the problem of semantic information handling (particularly in biological data streams, which inundate today's biological research). Therefore, it will be our duty to try and to find out the proper answers to the questions that computational biology people are feeling too shy to ask.

## 2  Semantic information: what does it mean?

I am not a biologist, so I do not possess the terminology commonly used by biological scientists. My experience comes from the fields of Remote visual inspection and Homeland security. Despite all their differences, many things are common to computational biology and visual signal analysis. First, the problem of Big Data deluge. High-resolution digital images are entering a visual system at a rate of 25 frames per second and huge amounts of image data (in each frame) have to be processed and analyzed in order to understand what is going on in the observable world. No kind of image processing can be applied unless you understand the semantic content of the image. That is the second thing that we have in common – none knows what the term "semantic content" means, how it has to be defined and how to use it properly.

Bearing in mind the mentioned above commonalities I will try to share with you my recent findings in the domain of image information content processing. Obviously, my explanations would be expressed in an image processing language. Nevertheless, I hope, this will not preclude the biologists from understanding the subject of this discourse.

For the sake of time and space saving, I will provide here only a short excerpt from my once published (and mostly unknown) papers. Interested readers are invited to visit my website (http://www.vidia-mant.info ), where more of such papers can be found and used for a further elaboration of issues relevant to our discourse.

Contrary to the widespread use of Shannon's Information Theory, my research relies on the Kolmogorov's definition of information, [4]. According to Kolmogorov, it can be expressed as such:

**"Information is a linguistic description of structures observable in a given data set".**

For the purposes of our discussion, digital image is a proper embodiment of a given data set. It is a two-dimensional array of a finite number of elements, each of which has a particular location and value.

These elements are regarded to as picture elements (also known as pels or pixels). It is taken for granted that an image is not a random collection of these picture elements, but, as a rule, the pixels are naturally grouped into specific assemblies called pixel clusters or structures. Pixels are grouped in these clusters due to the similarity in their physical properties (e.g., pixels' luminosity, color, brightness and as such). For that reason, I have proposed to call these structures **primary or physical data structures**.

In the eyes of an external observer, the primary data structures are further grouped and arranged into more larger and complex assemblies, which I propose to call secondary data structures. These secondary structures reflect human observer's view on the composition of the primary data structures, and therefore they could be called **meaningful or semantic data structures**. While formation of primary data structures is guided by objective (natural, physical) properties of the data, ensuing formation of secondary structures is a subjective process guided by human habits and customs, mutual agreements and conventions between and among members of the observing group.

As it was already stated above, **Description of structures observable in a data set should be called "Information".** Following the explained above subdivision of the structures discernible in a given image (in a given data set), two types of information must be distinguished – **Physical Information and Semantic Information**. They are both language-based descriptions; however, physical information can be described with a variety of languages (recall that mathematics is also a language), while semantic information can be described only by using natural human language.

I will drop the explanation how physical and semantic information are interrelated and interact among them. Although that is a very important topic, interested readers would have to go to the website and find there the relevant papers, which explain the topic in more details. Here, I will continue with an overview of the primary points that will facilitate our understanding of the issues of the further discourse.

Every information description is a top-down evolving coarse-to-fine hierarchy of descriptions representing various levels of description complexity (various levels of description details). Physical information hierarchy is located at the lowest level of the semantic hierarchy. The process of sensor data interpretation is reified as a process of physical information extraction from the input data, followed by an attempt to associate the input physical information with physical information already retained at the lowest level of a semantic hierarchy. If such association is achieved, the input physical information becomes related (via the physical information retained in the system) with a relevant linguistic term, with a word that places the physical information in the context of a phrase, which provides the semantic interpretation of it. In such a way, the input physical information becomes named with an appropriate linguistic label and framed into a suitable linguistic phrase (and further – in a story, a tale, a narrative), which provides the desired meaning for the input physical information. (Again, more details can be found on the website).

The segregation between physical and semantic information is the most essential insight about the nature of information. Its extraordinary importance should not be underestimated – the collapse of the Artificial Intelligence idea, the bankruptcy of the Machine Learning idea (semantic information is a mutual agreement between observers and thus cannot be learned, it must be only shared or granted), the failure of the everlasting attempts to derive semantic information from the omnipresent physical information (which is now a widespread practice in almost every scientific field) – all these are the result of the information dichotomy overlook.

Another important outcome from the semantic information definition is the comprehension about the form in which semantic information can be reified. That is, semantic information is always reified as a string of words, a piece of text, a narrative. That poses a problem when we think about further semantic information processing – computers are data processing machines, and how they can fit the task of text strings processing remains a problem and an unanswered question.

# 3  Implications for biological information processing

To alleviate the readers' comprehension of my text, I would like to propose a cross reference dictionary for some commonly used notions and terms.

The term "description" (used in my texts) is equivalent to the term "modeling" (frequently used in computational biology literature). Modeling usually means "describing mathematically a certain phenomenon". Even more often, the term is used in a "Data modeling" context. Which means: "the act of exploring data-oriented structures". That fits well my definition of physical information – "description of primary (physical) data structures" which can be done (besides other alternatives) in a mathematical language too. However, description of secondary (semantic) data structures requires exclusive use of natural language only. To preserve uniformity of the information definition, I prefer to use the term "description" in both cases.

The terms "primary structures" and "secondary structures" are frequently used in my and in computational biology texts. While "primary structures" have the same meaning in both cases, I am not sure that this is right for the case of secondary and higher order structures. In computational biology, secondary, 3-D and 4-dimensional structures are always seen as arrangements of raw data. They cannot be considered as semantic information entities.

As it was explained earlier, the semantic perception of the sensed data begins with physical information extraction from it. It must be emphasized that only physical information is being processed further in the semantic information-processing stream. All physical traits of the input data are lost at this stage. In the end, we understand the essence of an image ignoring its illumination conditions or color palette. The same is with speech perception – we understand the meaning of a phrase independent of its volume or gender voice differences.

The extracted physical information is associated then with the physical information retained at the lowest level of the semantic hierarchy. If a match is reached, the physical information becomes tied up with a linguistic term (a word on the higher level of the active semantic hierarchy), which is a part of a linguistic expression (a phrase, a saying). In such a way, it finds its place in a linguistic expression, which determines its meaning, its semantics. (Analogous to "comprehension from usage" or "understanding from action" forms of semantics disambiguating).

This physical data structures naming is in a close resemblance to the ontology-based annotation process. Ontologies are the most recent form of knowledge representation and are widely used in biomedical science enabling to turn data into knowledge. Despite of the resemblance, semantic information hierarchies and ontologies are strikingly different. From my definition of semantic information follows that 1) knowledge is memorized (retained in the system) information (and nothing else!), 2) semantic information is an observer's property, and 3) semantic information has nothing to do with data! That is, data is semantics devoid. So, the purpose of ontologies "to describe the semantics of data", [9], is misinterpreted. Computational biology tools developers have to pay more attention to this peculiarity.

The next point that deserves attention is: Semantics is a user's property, it cannot be derived from data, but it has to be provided from an external source of knowledge. If we speak about the first forms of life, about the single-celled microorganisms like bacteria (and their genome, where the acquired knowledge is actually installed) who was this "external source of knowledge" that shared his semantic information with the first members of the microbial "collective"?

I do not have a suitable answer for this question. But I am inclined to accept the Maynard Smith's guess that "In biology, the coder is natural selection", [10]. That is, natural selection had created the first reference knowledge base (the first semantic information description, the first memorized narrative) in the microbial gene in relation to which signal understanding and interpretation was subsequently worked up, thus paving the way for the first forms of cognitive behavior. Further editing and adaptive changes have enlarged and extended these initial lifetime stories, thus, producing the semantic information knowledge base conserved in the early genomes.

"It is in this sense that genomes have intentionality. Intelligent design and natural selection produce similar results. One justification for this view is that programs designed by humans to produce a result are similar to and may be indistinguishable from, programs generated by mindless selection", [10].

Today, these early conserved information descriptions serve the biologists (that are busy with gene texts deciphering) as prototype references for particular data sequences localization and annotation.

# 4  Conclusions

This paper is my first attempt to bring into attention of biological scientists my improved version of the information definition. It is not a thorough explanation of the concept; it is just a sketched introduction to the basic principles of information literacy. Inevitably, some of the "suckered cows" of the contemporary biological science should be and will be sacrificed on this occasion. Shannon's Information Theory, certainly, will be the first to be repudiated. Then the time of Biological Information will come.

Taking into account the proposed definitions of what should be called information the term "biological information" (as well as "economical information" or "astronomical information") has to be used with appropriate reservations. Because of the hidden and not so well understood duality of the term "information". Because of the two composing constituents concealed in the term – "physical information" and "semantic information" – which have to be treated carefully and consciously (as it is explained in this paper).

At this time, the science in general and computational biology in particular are undergoing a paradigm shift in their research approaches – from data-processing (computational) paradigm to information-processing (cognitive) paradigm. In image processing we witnessed that about a decade earlier – an attempt to shift from Computer Vision ("computer" means "computational") to Cognitive Vision design enterprises. However, the lack of understanding about the peculiarities of semantic information has impeded this process in image processing. For the same reasons, it does not take off in other branches of science and engineering. In spite of this, the transition from data-processing (computational) way of thinking to information-processing (cognitive) mode of operation is becoming more and more pervasive: cognitive science is replacing computational science, cognitive linguistics is replacing computational linguistics, and so on. The time is right to start the shift from computational biology to cognitive biology.

One of the obvious problems that arises in such a transition is as follows: We are accustomed to use computers in our everyday life. Computer is a data processing devise. Semantic information comes about as a text string. Therefore, semantic information processing must be treated as text strings processing. But that is not what our computers are supposed to do. There is an urgent need to invent a new generation of computers that will be capable to process natural language texts (which are the expression of semantic information).

I hope my humble elucidations will help Computational Biologists to find their right way in the present dangerous times of Big Data deluge.